# Data-driven prediction of room temperature density for multicomponent silicate-based glasses

[1]Kai Gong and [1*]Elsa Olivetti

[1]Department of Materials Science and Engineering, Massachusetts Institute of Technology, Cambridge, MA 02139, USA.

*Correspondence: Elsa A. Olivetti, Department of Materials Science and Engineering, MIT, Cambridge 02139, USA. Email: elsao@mit.edu

**Abstract**

Density is one of the most commonly measured or estimated materials properties, especially for glasses and melts that are of significant interest to many fields, including metallurgy, geology, materials science and sustainable cements. Here, two types of machine learning (ML) models (i.e., random forest (RF) and artificial neural network (ANN)) have been developed to predict the room-temperature density of glasses in the compositional space of CaO-MgO-$Al_2O_3$-$SiO_2$-$TiO_2$-FeO-$Fe_2O_3$-$Na_2O$-$K_2O$-MnO (CMASTFNKM), based on ~2100 data points mined from ~140 literature studies. The results show that the RF and ANN models give accurate prediction of glass density with $R^2$ values, RMSE, and MAPE of ~0.96-0.98, ~0.02-0.03 g/cm$^3$ and ~0.59-0.79%, respectively, for the 15% testing set, which are more accurate compared with empirical density models based on ionic packing ratio (with $R^2$ values, RMSE, and MAPE of ~0.28-0.91, ~0.05-0.15 g/cm$^3$, and ~1.40-4.61%, respectively). Furthermore, glass density is shown to be a reliable reactivity indicator for a range of CaO-$Al_2O_3$-$SiO_2$ (CAS) and volcanic glasses due to its strong correlation ($R^2$ values above ~0.90) with the average metal-oxygen dissociation energy (a structural descriptor) of these glasses. Analysis of the predicted density-composition relationships from these models (for selected compositional subspaces) suggests that the ANN model exhibits a certain level of transferability (i.e., ability to extrapolate to compositional space not (or less) covered in the database) and captures known features including the mixed alkaline earth effects for $(CaO-MgO)_{0.5}$-$(Al_2O_3-SiO_2)_{0.5}$ glasses.

# 1 Introduction

Silicate-based multicomponent glasses are ubiquitous in many important engineering applications and natural processes, including smartphone screens, optical fibers, building façades, ceramics, sustainable cements, metallurgic processes, volcanic activities and nuclear waste encapsulation [1-4]. One of the most important fundamental properties of interest for these glasses and their associated applications is density, which is often needed for the calculation of other important glass properties such as thermal conductivity, refractive index, and elastic and optical properties [5-8]. Hence, calculating and predicting the density of glasses and melt from their chemical composition has been an area of interest in many fields [5, 9-15], especially considering the high cost of density measurement at high temperatures [15].

One important direction for predicting composition-density relationships rests on empirical models. These include early density models based on partial density constant/factor of individual oxide component [16], the empirical polynomial model developed by Fluegel [10] based on silicate glass density data from SciGlass© and density models based on ionic packing ratio [5, 7, 13, 17, 18] and partial molar volume [14, 15]. These empirical models give density predictions with errors generally less than 10% [6]. Alternatively, data-driven machine learning (ML) models that capture the hidden trends in the composition-property relationships have also been employed for density prediction [6]. For example, Gaafar et al. developed an artificial neural network (ANN) model to predict the density and mechanical properties of tellurite glasses from their chemical composition [19]. Recently, Deng employed several ML models (including ANN and random forest (RF)) to predict the density and elastic properties of a wide range of industrial glasses based on a large data set collected at Corning Incorporated [20]. More recently, Ahmmad et al. also developed ANN and RF models to predict glass density for oxide [6] and fluoride glasses [21]. Another powerful ML technique, i.e., Gaussian process regression (GPR), has recently been used to predict glass density along with a range of other properties (e.g., Young's modulus, hardness, and thermal expansion coefficient) for a wide range of oxide glasses [22]. The density data used to train ML models in the aforementioned studies is mostly experimentally determined, whereas a recent study [23] employed high-throughput atomistic simulations to estimate the density and elastic moduli of silicate-based glasses, which were then used to train the ML model (i.e., least absolute shrinkage and selection operator with a gradient boost machine). In addition to density, ML techniques have also been increasingly used to predict other glass properties, including elastic properties, glass transition temperature, coefficient of thermal expansion, liquidus temperature, and refractive index, as has been briefly summarized in ref. [24].

Synthesizing across the aforementioned studies [5, 7, 10, 14, 15, 17-20, 21-24] reveals that there is still a lack of ML studies focusing on density prediction for silicate-based glasses that are relevant to cement and concrete applications. For example, most of these existing models are built based on glasses containing oxide components (e.g., $B_2O_3$, SrO, $La_2O_3$, $Eu_2O_3$ or $TeO_2$)[5, 10, 18-20, 23] that are not often present in the glassy raw materials commonly used for the production of modern concrete. These glassy (or amorphous) raw materials are often industrial by-products (e.g., blast furnace slags from steel production, fly ashes from coal-fired power plants, and waste glasses) or naturally derived materials (e.g., volcanic ashes and calcined clays), which have been used in concrete production to partially replace Portland cement (PC) [25-27] and would therefore lower the $CO_2$ burden associated with the use of PC (the dominant cement product in the current market). Partial replacement of PC (i.e., blended cement) has been identified as a major strategy to decarbonize the current concrete industry, which is responsible for 8-9% of the global anthropogenic $CO_2$ emissions [28].

These low-$CO_2$ raw materials can also be used to synthesize alkali-activated materials (AAMs), which boast great potential to lower the $CO_2$ footprint of concrete materials [29]. The main reactive components across these raw materials are often rich in glassy aluminosilicate phases with various levels of CaO, MgO, $Fe_2O_3$, FeO, MnO, $TiO_2$, $Na_2O$, and $K_2O$, depending on the sources [30-32]. This inherent chemical variability leads to differences in their reactivity as well as the engineering properties of the resulting cement and concrete products, as has been briefly discussed in ref. [33]. Therefore, establishing composition-structure-property relationships for silicate-based glasses is critical to their applications in blended cements and AAMs [33, 34], in addition to being a grand challenge for the glass community [35].

As part of the growing efforts to predict glass properties from their chemical composition, this investigation aims to develop ML models to estimate the room-temperature density of glasses in the compositional space of CaO-MgO-$Al_2O_3$-$SiO_2$-$TiO_2$-FeO-$Fe_2O_3$-$Na_2O$-$K_2O$-MnO (CMASTFNKM), relevant to blended cement and AAM applications. Specifically, we built an ensemble of RF and ANN models based on extracting ~2100 room-temperature glass density data from ~140 literature studies. We compared the performance of these ML models with several empirical density models based on ionic packing ratio, including linear and second-order polynomial models introduced in this study. Furthermore, we evaluated the potential use of density as a reactivity indicator for synthetic CaO-$Al_2O_3$-$SiO_2$ (CAS) glasses and natural volcanic glasses from three high-quality experimental studies [36-38]. Finally, we demonstrated how the ML density models can be used to explore composition-density relationships for CAS,

MgO-Al$_2$O$_3$-SiO$_2$ (MAS), and CaO-MgO-Al$_2$O$_3$-SiO$_2$ (CMAS) glasses, especially in the compositional space not (or less) covered in the original database.

## 2 Data and Methods

### 2.1 Data curation

We extracted ~2100 room-temperature glass density records along with the corresponding chemical composition from ~140 literature studies. The database is given in the Excel spreadsheet in Supplementary Material A. These density values are mostly experimentally measured, except for one study where 133 density values for binary Na$_2$O-SiO$_2$, K$_2$O-SiO$_2$, CaO-SiO$_2$ and Al$_2$O$_3$-SiO$_2$ glasses and ternary Na$_2$O-K$_2$O-SiO$_2$, Na$_2$O-CaO-SiO$_2$, Na$_2$O-Al$_2$O$_3$-SiO$_2$, K$_2$O-CaO-SiO$_2$, and CAS glasses have been estimated using force field molecular dynamics (MD) simulations [23]. Also, note that some minor oxide components that may have been present in the glasses have not been included in the database (e.g., SO$_3$ in blast furnace slags [31, 32]). Figure 1 shows the histograms of the molar oxide compositions and densities for all the glasses in the database, with their statistics given in Table S1 of Supplementary Material B. It is clear from Figure 1 and Table S1 that the glasses in this database cover a wide range of compositional space (i.e., 0-100% SiO$_2$, 0-60% Al$_2$O$_3$, 0-70% CaO, 0-50% MgO, 0-25% FeO, 0-20% Fe$_2$O$_3$, 0-40% MnO, 0-50% Na$_2$O, 0-60% K$_2$O, and 0-55% TiO$_2$) and density values (~2.2-3.3 g/cm$^3$). However, there are significantly fewer glasses in the database containing certain oxide components (as compared to others), especially MnO and FeO, and to a lesser extent, Fe$_2$O$_3$ and TiO$_2$.

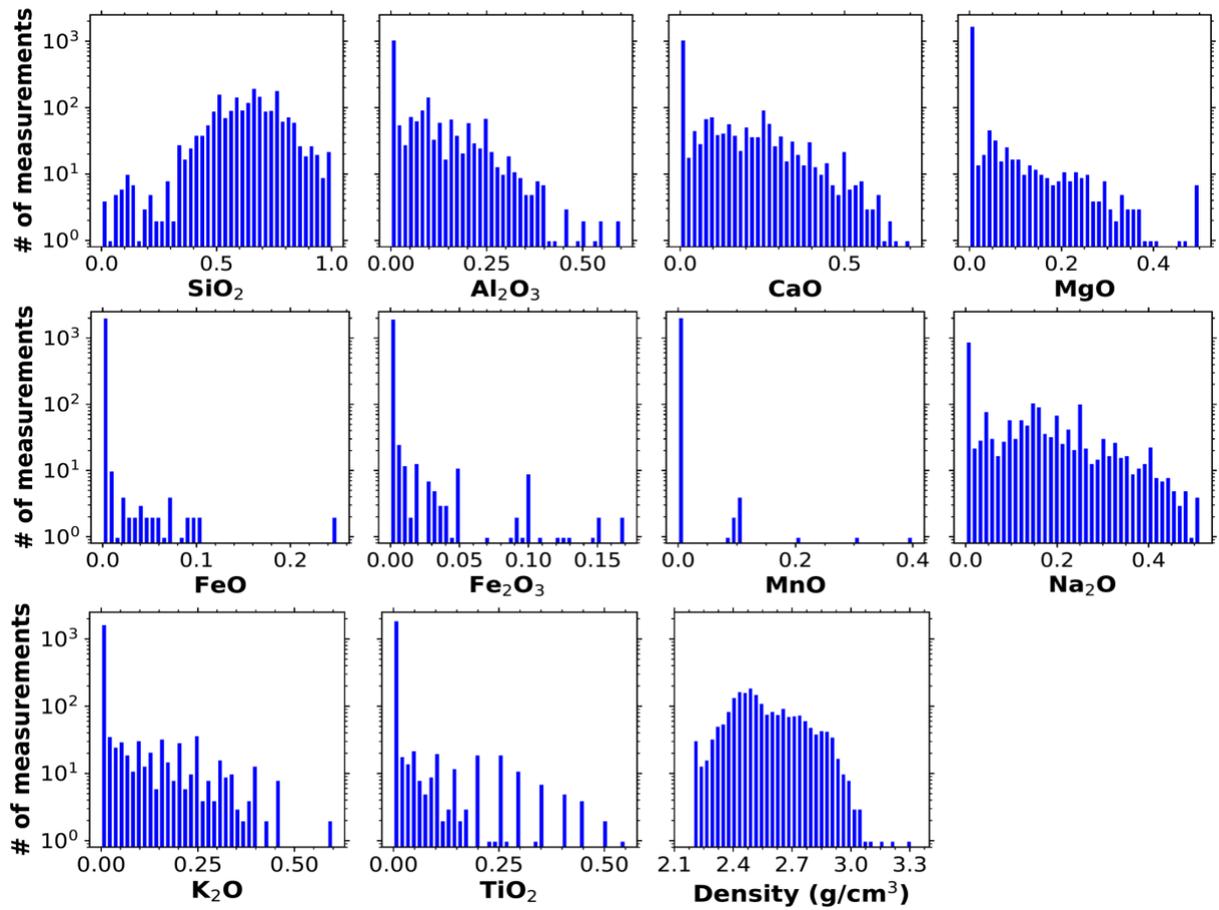

Figure 1. Histograms of the molar oxide content and density values of all the glasses in the mined database, as given in Supplementary Material A.

2.2 Machine learning models

As mentioned in the Introduction, different empirical models have been developed in the literature to predict glass density. In this investigation, we employed two types of ML algorithms to predict the room temperature densities of CMASTFNKM glasses as a function of their chemical compositions, namely ANN and RF, both of which have been used in the glass community to predict glass properties [6, 19-21]. ANN modeling works by propagating raw information from an input layer (i.e., chemical compositions of glasses here) through the hidden neurons in between (where the raw information is processed) all the way to the final output layer to generate a prediction (i.e., glass density). Each hidden neuron can be mathematically described using Eq. (1) below:

$$y(t) = F(\sum_{i=0}^{m} w_i(t) \cdot x_i(t) + b) \qquad (1)$$

where, $y(t)$ represents the output value at a discrete time $t$ from each neuron, $x_i(t)$ is the $i^{th}$ input value from the neurons in the previous layer at the discrete time $t$, $w_i(t)$ and $b$ are the weight value and bias applied to the input values, respectively, and $F$ is the activation function that transforms the weighted sum of the inputs to the output target.

As the default recommendation in modern ANN [39], the rectified linear activation unit (ReLU; $F(X) = \max(0, X)$) has been adopted here as the activation function. Single-layer ANN is used here as one hidden layer is often sufficient for most problems, according to Heaton [40]. Furthermore, we used the Limited-memory Broyden-Fletcher-Goldfarb-Shanno (L-BFGS) algorithm to iteratively update weights and bias to minimize the loss function due to its fast convergence and superior performance for small datasets [41, 42]. Finally, we have optimized three hyperparameters for the ANN models, namely (i) the number of training epochs, (ii) the number of neurons in the hidden layer, and (iii) the L² regularization parameter (i.e., alpha) used to reduce overfitting. RF modeling works by using an ensemble of decision trees (generated using a bootstrap aggregating technique) to make a prediction, where the final model prediction is averaged over the outputs produced by all the trees. For the RF models, we have also optimized three important hyperparameters, namely (i) the number of trees (n_estimators), (ii) the number of features to consider at each split (max_features), and (iii) the minimum number of samples at a leaf node (min_samples_leaf).

Based on the mined database in Section 2.1, we have built an ensemble of ANN and RF models, following similar procedures adopted in our recent study on quartz dissolution rate prediction [43], as also briefly outlined here. First, we performed a shuffled and stratified split of the data, consisting of nine input features (i.e., molar content of all oxides except for $SiO_2$ since all ten oxide content add up to one) and one output target (i.e., density), into a training and a testing set with 85% and 15% of the data, respectively. This stratification based on density value ensures that both the training and testing sets contain the same proportion of data within each of the ten equally spaced density ranges. The training set (1761 measurements) was used to build the model (i.e., to learn the correlation between the input features and the output target) and determine the optimal hyperparameters. Specifically, we performed a stratified ten-fold cross-validation on the training set, which has been partitioned into ten equally sized folds. At each training-validation iteration, nine folds of the data were used for training while the remaining fold was held for validation. A grid search method was employed together with the ten-fold cross-validation to determine the optimal hyperparameters, which were then used to make predictions for the unexposed 15% testing set. We then repeated the whole process twenty times by using a different random state each time during the initial 85%-15% training-testing split, which enables the robustness of the ML models to be evaluated. More details on the model construction process, including the optimized hyperparameters, are given in Section 2 of Supplementary Material B. Three different error metrics, i.e., (i) mean absolute percentage error (MAPE), (ii) root mean square error (RMSE), and (iii) coefficient of the determinant ($R^2$), have been calculated to evaluate model performance, with the calculation details given in Section S3 of Supplementary Material B. All the ML modeling was implemented in Scikit-learn and executed in Python coding [42].

### 2.3 Force field molecular dynamics (MD) simulations

Force field MD simulations have been performed to generate detailed atomic structural representations for six CAS glasses with very different chemical compositions (see Table 1) for a case study in Section 3.3.1. The impact of glass composition on glass reactivity for the six CAS glasses has been experimentally studied in ref. [38]. The purpose of generating detailed structural representations for these glasses is to use the obtained structural information to calculate the so-called average metal-oxygen dissociation energy (AMODE) parameter, which gives an overall estimation of the average energy required to break all the metal-oxygen bonds in a glass. The resulting AMODE values of the six CAS glasses will then be compared with their densities in Section 3.3.1 to evaluate the feasibility of using glass density as an indicator of CAS glass reactivity. The AMODE parameter is defined according to Eq. (2) [33, 34]:

$$AMODE = \frac{\sum N_M \cdot CN_M \cdot E_{M-O}}{\sum N_M} \quad (2)$$

where $N_M$ is the number of metal cation $M$ ($M$ = Ca, Si, Al) in the CAS glass, $E_{M-O}$ and $CN_M$ represent the average energy required to break a single metal-oxygen bond and the average coordination number (CN) (i.e., the number of metal-oxygen bonds formed within the first coordination shell), respectively. The $E_{M-O}$ parameters are obtained from ref. [44] (as also given in Table S4 of Supplementary Material B), while the $CN_M$ parameters can be calculated from the atomic structures generated using MD simulations, as described below.

Table 1. Chemical composition of the six CaO-$Al_2O_3$-$SiO_2$ (CAS) glasses studied here (obtained from ref. [38]).

| # ID | Type | Composition (mole %) | | | Density (g/cm³) at different temperatures (K) | | |
|---|---|---|---|---|---|---|---|
| | | CaO | $Al_2O_3$ | $SiO_2$ | !300 | *2000 | *5000 |
| 1 | Slag | 48.4 | 11.8 | 39.9 | 2.92 | 2.55 | 2.35 |
| 2 | Slag | 40.2 | 9.8 | 50.0 | 2.81 | 2.46 | 2.26 |
| 3 | Fly ash | 24.8 | 25.3 | 49.9 | 2.69 | 2.40 | 2.20 |
| 4 | Fly ash | 17.9 | 17.7 | 64.5 | 2.56 | 2.30 | 2.10 |
| 5 | Natural pozzolan | 7.3 | 7.1 | 85.6 | 2.31 | 2.16 | 1.96 |
| 6 | Silica fume | 0.0 | 0.0 | 100.0 | 2.20 | 2.08 | 1.88 |

!Experimental density values at 300 K is obtained from ref. [45]; *High temperature densities are estimated numerically, as detailed in Section 5 of Supplementary Material B.

Atomic structural representations for the six CAS glasses in Table 1 have been generated following a commonly used "melt-and-quench" method in MD simulations [46, 47], as has been given in detail in several previous studies [33, 34, 48]. Briefly, we started by melting the structure in a simulation box containing about 4000 atoms (with similar chemical compositions as the experimental data, Table 1) at a temperature of 5000 K, and then progressively quenching the structure to 2000 K (over 1.5 ns) and then to 300 K (over 2 ns). At each target temperature (i.e., 5000, 2000, and 300 K), the structure has been equilibrated by 0.5–1 ns. The canonical *NVT* ensemble with the Nosé Hoover thermostat and a time step of 1 fs has been adopted for all the simulation steps. Prior to each equilibration step, the density of the simulation cell has been adjusted to either an experimentally measured value (available at 300 K from the original study [38]) or a numerically estimated value (at 2000 and 5000 K, see Table 1 and the calculation details in Supplementary Material B). With the empirical density models and the data-driven ML models developed in this article, the density of CAS glasses (and other types of CMASTFNKM glasses) at room temperature can also be readily estimated for future use in MD simulations.

All the MD simulations have been performed using the ATK-Forcefield module in the QuantumATK NanoLab software package [49, 50] and the Pedone force field [51], which has been developed to cover complex silicate crystals, melts, and glasses in the compositional space of CMASTFNKM. The mathematical description of the Pedone force field is given by Eq. (3) [51]:

$$U_{ij}(r_{ij}) = \frac{z_i z_j}{r_{ij}} + D_{ij}\left[\left\{1 - e^{-a_{ij}(r_{ij}-r_0)}\right\}^2 - 1\right] + \frac{C_{ij}}{r_{ij}^{12}} \qquad (3)$$

where $z_i$ and $z_j$ are the effective charges associated with atom *i* and *j*, respectively, $r_{ij}$ is the interatomic distance between atom pair *i-j*, and $r_0$ is the equilibrium bond distance. $D_{ij}$, $a_{ij}$, and $C_{ij}$ are empirical parameters developed in ref. [51] by fitting elastic constants and structural parameters of various binary oxides using free energy minimization. The three terms in Eq. (3) represent the long-range Coulomb interaction, the short-range Morse function, and a repulsive contribution (required to model interactions at high temperature and pressure), respectively [51]. The force field parameters adopted here are summarized in Table 2. For each glass, 500 structural snapshots have been extracted from the last 500 ps of equilibration at 300 K and used to calculate the CN values, which were then used in Eq. (2) to estimate the corresponding AMODE parameter of the glass. CN values of Si, Al, and Ca atoms were calculated using a cut-off distance of 2.1, 2.4, and 3.1 Å, respectively, which were determined from the first minima of the corresponding partial radial distribution functions, similar to previous studies [33, 34, 48, 52-54]. The AMODE parameter will be used in Section 3.3 to evaluate the feasibility of using predicted glass density values (from the different density models) as an indicator of glass reactivity.

Table 2. Summary of force field parameters used for the Pedone force field[51].

| Atom | $z$ (e) | Metal-oxygen pairs | $D$ (eV) | $a$ (Å$^{-1}$) | $r_0$ (Å) | $C$ (Å$^{12}$ eV) | $r_{cut}$ (Å) for the three terms in Eq. (3) | | |
|---|---|---|---|---|---|---|---|---|---|
| | | | | | | | 1st | 2nd | 3rd |
| Si | 2.4 | Si-O | 0.340554 | 2.0067 | 2.1 | 1 | 12 | 7.5 | 7.5 |
| Al | 1.8 | Al-O | 0.361581 | 1.90044 | 2.16482 | 0.9 | 12 | 7.5 | 7.5 |
| Ca | 1.2 | Ca-O | 0.030211 | 2.24133 | 2.92325 | 5 | 12 | 7.5 | 7.5 |
| O | -1.2 | O-O | 0.042395 | 1.37932 | 3.6187 | 22 | 12 | 7.5 | 7.5 |

## 3  Results & Discussion

We develop two new empirical density models based on the ionic packing ratio and two data-driven ML models (i.e., RF and ANN) based on the mined database in Section 2.1 and compare their performance for predicting room-temperature glass density. We also evaluate the potential use of density as a reactivity indicator for synthetic CAS and natural volcanic glasses from three high-quality experimental studies [36-38]. We then compare the ability of these models to predict composition-density relationships for selected compositional spaces.

3.1 Empirical density models based on ionic packing ratio

Empirical models based on ionic packing ratio have been widely used for glass density prediction.[5, 7, 13, 17, 18] The ionic packing ratio $V_p$ is defined using Eq. (4) [55].

$$V_p = \rho \frac{\sum V_i \cdot x_i}{\sum M_i \cdot x_i} \tag{4}$$

where $V_i$, $M_i$ and $x_i$ are ionic packing factor, molar weight, and molar fraction of the $i^{th}$ oxide component, respectively. The ionic packing factor of the individual oxide component $V_i$ is calculated using Eq. (5) [5].

$$V_i = \frac{4}{3}\pi \cdot N_A \cdot (X \cdot r_M^3 + Y \cdot r_O^3) \tag{5}$$

where $N_A$ is the Avogadro's number (mol$^{-1}$), $r_M$ and $r_O$ are the ionic radius of the metal and oxygen, respectively. Here, we calculated $V_i$ value for each oxide component using Pauling's effective ionic radii [56], which are given in Table S5 of Supplementary Material B. The density of glasses, $\rho$, is then estimated using the $V_i$ values and Eq. (6).

$$\rho = V_p \frac{\sum M_i \cdot x_i}{\sum V_i \cdot x_i} \tag{6}$$

where the ionic packing ratio $V_p$ can be approximately taken as a constant, e.g., 0.53, 0.56 or 0.57, according to several previous studies [5, 7, 13]. The predictive performance of Eq. (6) with constant $V_p$ of 0.53, 0.56, and 0.57 is presented in Fig. S2 of Supplementary Material B, with the corresponding error metrics summarized in Table 3. It is clear from Table 3 and Fig. S2 that a constant $V_p$ of 0.56 and 0.57 exhibit better predictive performance (e.g., higher $R^2$ values and lower RMSE and MAPE) than a constant $V_p$ of 0.53. This observation is consistent with two previous studies on CAS [7] and CMAS glasses [13], where an ionic packing ratio $V_p$ of 0.56-0.67 was also seen to give better density predictions than 0.53. See more discussion of Fig. S2 in Section 7 of Supplementary Material B.

Table 3. A comparison of the predictive performance of the density models based on Eq. (6) with a constant ionic packing ratio $V_p$. RMSE = root mean square error; MAPE = mean absolute percentage error; $R^2$ = coefficient of determination. The calculation details for RMSE, MAPE, and $R^2$ are given in Section 3 of Supplementary Material B.

| Density model based on Eq. (6) | RMSE (g/cm$^3$) | MAPE | $R^2$ value |
|---|---|---|---|
| $V_p = 0.53$ | 0.146 | 4.607% | 0.276 |
| $V_p = 0.56$ | 0.086 | 2.777% | 0.748 |
| $V_p = 0.57$ | 0.106 | 3.610% | 0.615 |

However, as shown in Figure 2a, the ionic packing ratio $V_p$ is dependent on glass density (hence glass composition), varying considerably between 0.50 and 0.64 for the glasses studied here, as opposed to adopting a constant value. The $V_p$ parameter appears to be positively and linearly correlated with the measured glass density (an $R^2$ value of ~0.81 is achieved for linear regression, Figure 2a), with a higher $V_p$ associated with a higher glass density. This positive linear correlation between density and ionic packing ratio has also been observed in a previous study on Al$_2$O$_3$–SiO$_2$ binary glasses [57].

Figure 2b shows that the correlation between measured glass densities and $\sum(M_i \cdot x_i)/\sum(V_i \cdot x_i)$ is better captured by a linear and a second-order polynomial functions not passing through the origin (0,0) compared with Eq. (6). The linear and second-order polynomial functions for the best fit of the data in Figure 2b are given by Eqs. (7) and (8), respectively.

$$\rho = 0.866 \cdot \frac{\sum M_i \cdot x_i}{\sum V_i \cdot x_i} - 1.430 \qquad (7)$$

$$\rho = -0.525 \cdot \left(\frac{\sum M_i \cdot x_i}{\sum V_i \cdot x_i}\right)^2 + 5.819 \cdot \frac{\sum M_i \cdot x_i}{\sum V_i \cdot x_i} - 13.082 \qquad (8)$$

The predicted density values according to Eqs. (7) and (8) are compared with the corresponding measured densities in Figures 2c and 2d, respectively, which show that both the linear model (Eq. (7)) and the second-order polynomial model (Eq. (8)) exhibit better predictive performance than Eq. (6) with a constant $V_p$. For example, the linear and polynomial models have reduced (increased) the MAPE ($R^2$ values) from ~2.78-4.61% (~0.28-0.75) to ~1.40-1.57% (~0.88-0.91). Furthermore, comparing Figures 2c and 2d reveals that the polynomial model gives slightly better predictions than the linear model, which can be attributed to the better capturing of the data in Figure 2b at both the low-density (e.g., < 2.3 g/cm³) and high-density (e.g., > 2.9 g/cm³) regions for the polynomial model.

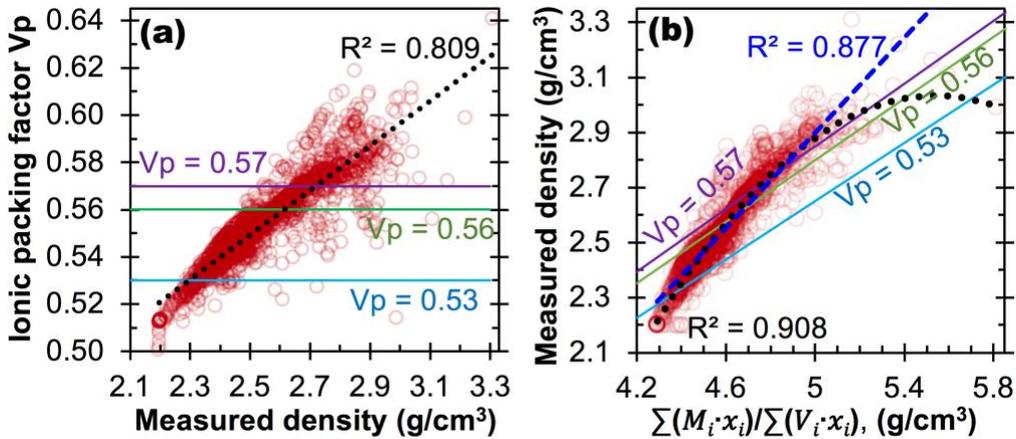

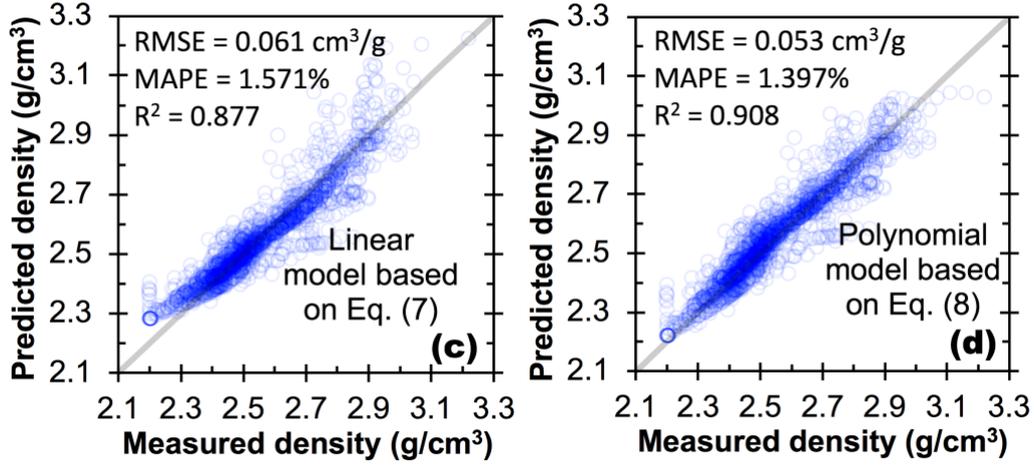

Figure 2. (a) Correlation between the measured glass density and the ionic packing ratio $V_p$ calculated using Eq. (4). The $R^2$ value in (a) is obtained by linear regression of the data. (b) Correlation between the measured glass density and $\sum(M_i \cdot x_i)/\sum(V_i \cdot x_i)$ for all the glasses in the mined database in Supplementary Material A, with the $R^2$ values for the fit of the linear (black dotted line) and second-order polynomial (blue dashed line) functions given in the figure. The solid lines in (b) represent Eq. (6) with constant $V_p$ of 0.53, 0.56, and 0.57. (c) and (d) show the predictive performance of the linear (Eq. (7)) and polynomial (Eq. (8)) models, respectively, with the corresponding RMSE, MAPE, and $R^2$ values given in the figures. The grey lines in (c) and (d) represent the line of equality.

3.2  Machine learning models

In this section, we evaluate the performance of the two types of ML models (i.e., RF and ANN) as described in Section 2.2 to predict CMASTFNKM glass density.

*3.2.1  Random forest model*

Figure 3 compares the measured glass densities with the predicted values from a typical RF model (see Section 2.2 and Supplementary Material B for details on model construction), which shows that the RF model gives accurate predictions of room temperature densities for the silicate-based glasses investigated here, with only a small proportion of predictions (< 3%) that are over 2% away from the measured density values. We see that a MAPE of 0.753%, an $R^2$ value of 0.970 and an RMSE of 0.030 g/cm$^3$ are achieved for the 15% testing set (311 samples), which has not been exposed to the RF model trained and validated on the 85% training set. Table 4 summarizes the error matrices for both the training and testing sets, averaged over twenty RF models, each developed using a different random state during the initial training-testing split. The relatively small standard deviations suggest that the analysis is robust. It is clear that the RF models exhibit noticeably larger error metrics for the testing set than the training set, suggesting that there is some degree of overfitting. Nevertheless, a comparison of these error metrics for the testing set with those achieved using the empirical density models based on ionic packing ratio (see Table 3 and Figures 2c and 2d) shows that the data-driven RF model exhibits superior predictive performance, even compared with the best-performing polynomial model in Figure 2d, which has a MAPE, RMSE, and $R^2$ value of 1.397%, 0.053 g/cm$^3$, and 0.908, respectively.

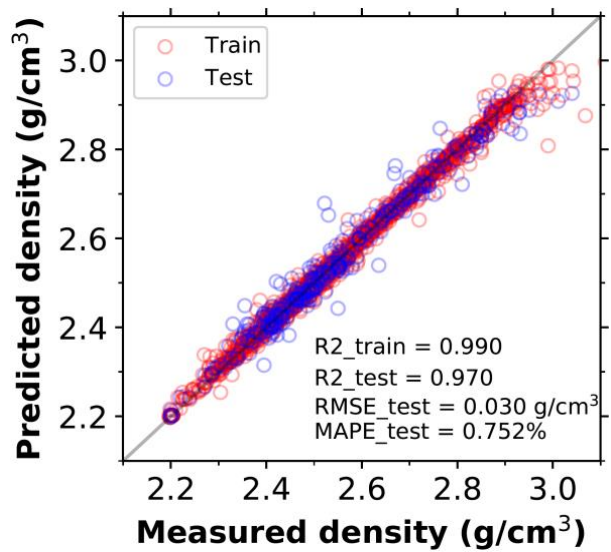

Figure 3. Comparison of the measured glass densities with the predicted values from a typical RF model. Selected error metrics (i.e., MAPE, $R^2$ and RMSE) are also shown in the figure,

with the calculation details given in Section 3 of Supplementary Material B. The error metric for training (i.e., R2_train) has been calculated using all the data in the 85% training set, which includes both the training and validation folds from the ten-fold cross-validation process. The solid grey line represents the line of equality.

Table 4. Summary of the error metrics (i.e., $R^2$ value, RMSE, and MAPE) for the RF and ANN models. Each average value and standard deviation (Stdev) are obtained from twenty model predictions, with each model trained independently using a different random state for the initial 85%-15% training-testing split. The error metrics for training have been calculated using all the data in the 85% training set, which includes both the training and validation folds from the ten-fold cross-validation process.

| Model type | Type of error metric | Model performance | | | |
|---|---|---|---|---|---|
| | | Train | | Test | |
| | | Average | Stdev | Average | Stdev |
| RF | MAPE (%) | 0.338 | 0.010 | 0.787 | 0.058 |
| | RMSE (g/cm$^3$) | 0.016 | 0.001 | 0.034 | 0.006 |
| | $R^2$ value | 0.992 | 0.001 | 0.958 | 0.014 |
| ANN | MAPE (%) | 0.505 | 0.019 | 0.592 | 0.039 |
| | RMSE (g/cm$^3$) | 0.021 | 0.001 | 0.024 | 0.003 |
| | $R^2$ value | 0.986 | 0.001 | 0.979 | 0.005 |

### 3.2.2 Artificial neural network model

We evaluated the predictive performance of single-layer ANN models, with results from a typical model presented in Figure 4. The single-layer ANN model also exhibits accurate predictions of glass densities, with a MAPE, $R^2$ value, and RMSE of 0.568%, 0.982, and 0.023 g/cm$^3$ for the 15% testing test, respectively. Comparing these error metrics with those achieved by the RF model (see Figure 3) reveals that the predictive performance of the ANN model is slightly better. The summary of error metrics in Table 4 shows that the differences in prediction errors between the training and testing sets are smaller for the ANN model than for the RF model. For example, for the ANN models, the average MAPE and RMSE of testing is about 20% higher than that of training, whereas for the RF models, the average MAPE and RMSE of testing is about 120% higher than that of training. This suggests that the ANN model has slightly better transferability than the RF model for the glass density data studied here.

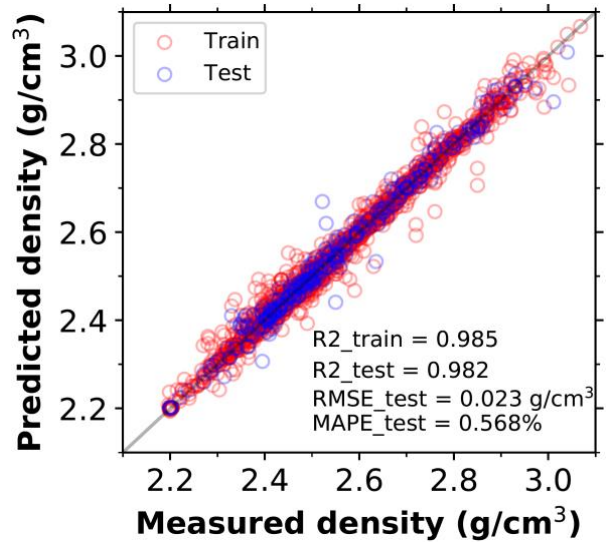

Figure 4. Comparison of the measured glass densities with the predicted values from an ANN model trained using the same 85%-15% training-testing split as the RF model in Figure 3. Selected error metrics (i.e., MAPE, $R^2$ and RMSE) are also shown in the figure, with the calculation details given in Section 3 of Supplementary Material B. The error metric for training (i.e., R2_train) has been calculated using all the data in the 85% training set, which includes both the training and validation folds from the ten-fold cross-validation process. The solid grey line represents the line of equality.

Compared with the empirical density models based on ionic packing ratio, including the linear and polynomial models introduced in this investigation (i.e., Eqs. (7) and (8) in Section 3.1), the two data-driven ML models (i.e., RF and ANN) give noticeably more accurate density predictions for the CMASTFNKM glasses covered here. The superior predictive performance of these ML models is attributed to their ability to extract the complex relationships between inputs (oxide composition) and output (density). However, these data-driven methods may have also captured noise (i.e., density variations not caused by chemical compositions) that may be present in the data. These include (i) the density differences caused by the use of different testing methods adopted by the different studies (see the references in Supplementary Material A), (ii) the differences between experiments and simulations (density data from one study [23] is from force field MD simulations), (iii) the presence of minor oxide components in some glasses (e.g., $SO_3$) that have not been considered in this study, and (iv) the potential presence of minor mineral phases or occurrence of phase segregation in some glasses.

3.3 Density as an indicator of glass reactivity?

Accurately predicting glass density from its chemical composition has several important implications. Glass density can be used to calculate other important glass properties, such as thermal conductivity, refractive index, and elastic and optical properties [5-7]. Second, estimating glass density from the chemical composition is relevant for glass modeling using MD simulations and the classical *NVT* ensemble [33, 34]. The reactivity of the glassy phases present in low-$CO_2$ raw materials (e.g., blast-furnace slag, fly ash, and volcanic ash) that have been used in blended cements and alternative low-$CO_2$ cements (e.g., alkali-activated materials) and their associated composition-properties [58] and composition-structure-properties [33, 34] relationships are critical to the use of these low-$CO_2$ raw materials in concrete for the purpose of reducing the $CO_2$ footprint of the cement and concrete industry [25]. Hence, developing parameters (based on either solely compositional information [58] or a combination of compositional and structural information [33, 34]) that are able to capture glass reactivity is of significant interest to the cement and concrete community. Such parameters are also of significant interest to the broader glass community, where important progress has been made recently [46, 59-61].

Here, we present a preliminary analysis of the potential of using density as an indicator of glass reactivity for glass compositions relevant to the cement and concrete community. This analysis has been motivated by the general observation in the literature that blast-furnace slags are often more reactive than fly ashes,[25] with the former (~2.8-3.0 g/cm$^3$) often having a higher density than the latter (generally lower than ~2.8 g/cm$^3$). A recent study on synthetic CAS glasses [37] (the main reactive component in many of these low-$CO_2$ raw materials) shows that the CAS glass with a slag composition has a higher reactivity and density (~2.93 g/cm$^3$) than CAS glasses relevant to fly ash (with a density of ~2.49-2.85 g/cm$^3$).

We evaluated the performance of glass density as a reactivity indicator based on three high-quality literature studies [36-38], which have experimentally investigated the impact of glass composition on glass reactivity under different environments. The first study [37], presented in Figure 5a, quantified the extent of reaction for eight synthetic CAS glasses (compositions given in Table 5) after 180 days of reaction in a mixture of portlandite, limestone, and sodium hydroxide using quantitative X-ray diffraction (XRD) analysis. Given that a larger surface area leads to faster reaction,[58] the extent of reaction data in Figure 5 has been normalized by the surface area of the synthetic glasses (with ~20% variation between the eight glasses), similar to a recent study [33]. Figure 5a shows positive and almost linear correlations between the extent of reaction and CAS glass densities estimated using different models, with a higher density generally associated with a greater extent of reaction (and hence a higher reactivity). The resulting $R^2$ values achieved for linear regression (0.93-0.97) are higher than that of the extent of depolymerization parameter used in the original article [37] ($R^2$ = ~0.70 [33]). A comparison of the three models (i.e., the ANN and RF models in Section 3.2 and the second-order polynomial model in Section 3.1) shows that the predicted densities from these models have a similar level of ability to capture the relative reactivity of the six CAS glasses.

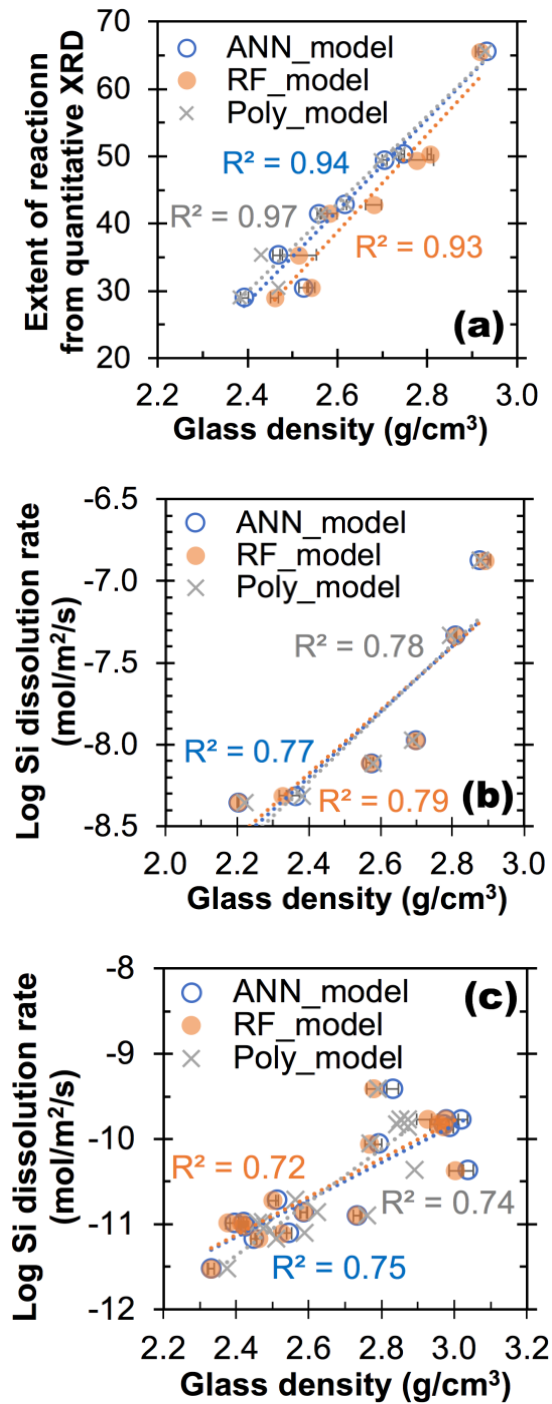

Figure 5. Correlation between predicted glass densities based on the polynomial, RF, and ANN models and reactivity data for the (a) eight synthetic $CaO-Al_2O_3-SiO_2$ (CAS) glasses in ref. [37] (b) six synthetic CAS glasses in ref. [38], and (c) seventeen volcanic glasses in ref. [36].

Table 5. Chemical compositions of the eight synthetic CAS glasses in ref. [37] and seventeen volcanic glasses in ref. [36].

| Ref. | ID | SiO$_2$ | Al$_2$O$_3$ | CaO | MgO | TiO$_2$ | FeO | Fe$_2$O$_3$ | MnO | Na$_2$O | K$_2$O |
|---|---|---|---|---|---|---|---|---|---|---|---|
| 37 | G1 | 0.840 | 0.106 | 0.054 | 0.000 | 0.000 | 0.000 | 0.000 | 0.000 | 0.000 | 0.000 |
|  | G2 | 0.771 | 0.173 | 0.056 | 0.000 | 0.000 | 0.000 | 0.000 | 0.000 | 0.000 | 0.000 |
|  | G3 | 0.706 | 0.241 | 0.054 | 0.000 | 0.000 | 0.000 | 0.000 | 0.000 | 0.000 | 0.000 |
|  | G4 | 0.660 | 0.175 | 0.165 | 0.000 | 0.000 | 0.000 | 0.000 | 0.000 | 0.000 | 0.000 |
|  | G5 | 0.655 | 0.103 | 0.242 | 0.000 | 0.000 | 0.000 | 0.000 | 0.000 | 0.000 | 0.000 |
|  | G6 | 0.547 | 0.169 | 0.284 | 0.000 | 0.000 | 0.000 | 0.000 | 0.000 | 0.000 | 0.000 |
|  | G7 | 0.457 | 0.246 | 0.296 | 0.000 | 0.000 | 0.000 | 0.000 | 0.000 | 0.000 | 0.000 |
|  | G8 | 0.358 | 0.093 | 0.550 | 0.000 | 0.000 | 0.000 | 0.000 | 0.000 | 0.000 | 0.000 |
| 36 | 1BT | 0.830 | 0.084 | 0.006 | 0.002 | 0.001 | 0.001 | 0.003 | 0.000 | 0.042 | 0.032 |
|  | 2O62 | 0.795 | 0.086 | 0.012 | 0.000 | 0.002 | 0.011 | 0.010 | 0.000 | 0.059 | 0.024 |
|  | 3A75 | 0.775 | 0.082 | 0.034 | 0.016 | 0.008 | 0.020 | 0.010 | 0.000 | 0.041 | 0.016 |
|  | 4H3B | 0.743 | 0.097 | 0.039 | 0.007 | 0.004 | 0.036 | 0.009 | 0.000 | 0.051 | 0.015 |
|  | 5HZ0 | 0.701 | 0.101 | 0.055 | 0.023 | 0.008 | 0.042 | 0.011 | 0.000 | 0.047 | 0.011 |
|  | 6H20 | 0.618 | 0.095 | 0.080 | 0.047 | 0.017 | 0.071 | 0.019 | 0.000 | 0.044 | 0.009 |
|  | 7GR | 0.552 | 0.087 | 0.115 | 0.090 | 0.021 | 0.093 | 0.011 | 0.000 | 0.029 | 0.004 |
|  | 8HEI | 0.577 | 0.109 | 0.085 | 0.043 | 0.019 | 0.074 | 0.016 | 0.000 | 0.065 | 0.012 |
|  | 9KRA | 0.541 | 0.086 | 0.119 | 0.093 | 0.016 | 0.105 | 0.013 | 0.000 | 0.025 | 0.002 |
|  | 10KAT | 0.530 | 0.084 | 0.110 | 0.080 | 0.038 | 0.104 | 0.017 | 0.000 | 0.030 | 0.006 |
|  | 11H1 | 0.786 | 0.090 | 0.023 | 0.001 | 0.002 | 0.021 | 0.005 | 0.000 | 0.053 | 0.019 |
|  | 12H3W | 0.783 | 0.091 | 0.025 | 0.002 | 0.002 | 0.022 | 0.005 | 0.000 | 0.053 | 0.018 |
|  | 13SLN | 0.719 | 0.097 | 0.036 | 0.018 | 0.011 | 0.040 | 0.009 | 0.000 | 0.051 | 0.019 |
|  | 14HZ1 | 0.622 | 0.100 | 0.080 | 0.050 | 0.017 | 0.053 | 0.027 | 0.000 | 0.041 | 0.009 |
|  | 15HZ3 | 0.598 | 0.099 | 0.086 | 0.057 | 0.021 | 0.069 | 0.026 | 0.000 | 0.037 | 0.008 |
|  | 16A61 | 0.557 | 0.083 | 0.107 | 0.077 | 0.023 | 0.100 | 0.021 | 0.000 | 0.029 | 0.004 |
|  | 17SS | 0.515 | 0.107 | 0.117 | 0.097 | 0.020 | 0.089 | 0.011 | 0.000 | 0.039 | 0.005 |

The second study,[38] presented in Figure 5b, also focused on synthetic CAS glasses, where the initial glass dissolution rate of six CAS glasses with very different chemical compositions (as given in Table 2) at pH 13 and 20 °C was investigated using inductively coupled plasma-optical emission spectrometry (ICP-OES). Figure 5b shows that a higher glass density is associated with a higher initial dissolution rate (based on Si release) for the six CAS glasses considered, consistent with the general trend observed in Figure 5a. Figure 5c compares glass density with dissolution rate data (based on Si release) collected on seventeen natural volcanic glasses with different chemical compositions (see Table 5) at pH 4, 20 °C, and far-from-equilibrium conditions, using ICP-OES (data obtained from ref. [38]). The comparison also reveals a positive correlation between glass density and the rate of dissolution for the seventeen volcanic glasses, with a higher density associated with a generally higher Si dissolution rate (hence a higher reactivity). Compared with the data in Figures 5a-b, the density-reactivity correlation in Figure 5c exhibits higher regional variations and lower $R^2$ values, which can be partially attributed to the more complex compositions of natural volcanic glasses compared with pure synthetic CAS glasses (see Tables 5 and 2). Comparing the glass density data from the three density models in Figure 5b-c also shows that the predictive performance of glass density is not affected by the type of density model used.

Overall, the results in Figure 5 suggest that glass density could be an indicator of glass reactivity, especially for the simpler CAS glasses. Intuitively, one would expect a lower reactivity for a denser (hence more compact) glass. To inform the underlying reason for the trends seen in Figure 5, we have plotted glass density as a function of the AMODE parameter for the glasses in the three above-mentioned studies [36-38] in Figure 6a-c, respectively. The AMODE values in Figures 6a and 6c are obtained from ref. [33] and ref. [34], respectively, whereas those in Figure 6b are calculated based on MD simulations performed here (see simulation details in Section 2.3). The CN results from the MD simulations for the six CAS glasses in Figures 5b and 6b are given in Table 6, which are then used in Eq. (2) of Section 2.3 to calculate the AMODE values. According to its definition, the AMODE parameter gives an overall estimation of the average energy required to break all the metal-oxygen bonds in a glass, and hence glasses with a higher AMODE value are expected to be harder to break/dissolve (i.e., being less reactive). Previously, the AMODE parameter has been shown to accurately capture the relative reactivity of C(M)AS[33] and volcanic glasses [34]. Here, the AMODE parameter is also seen to accurately capture the relative reactivity of the six CAS glasses as seen by the inverse correlation in Figure 7.

Table 6. Summary of the CN of the six synthetic CAS glasses in Table 2 based on MD simulations and the resulting AMODE value calculated using Eq. (2). The CN is determined at a cut-off distance of 2.1, 2.4, and 3.1 Å for Si-O, Al-O, and Ca-O pairs, respectively. These cut-off distances have been determined from the first minima of the corresponding partial radial distribution function, as commonly adopted in the glass modeling literature[33, 52-54, 62].

| # ID | Type | Coordination number (CN) | | | AMODE (kcal) |
|---|---|---|---|---|---|
| | | Ca-O | Al-O | Si-O | |
| 1 | Slag | 6.34 | 4.05 | 4.00 | 315.0 |
| 2 | Slag | 6.22 | 4.05 | 4.00 | 330.4 |
| 3 | Fly ash | 6.44 | 4.12 | 4.00 | 355.7 |
| 4 | Fly ash | 6.15 | 4.08 | 4.00 | 370.6 |
| 5 | Natural pozzolan | 5.53 | 4.04 | 4.00 | 398.8 |
| 6 | Silica fume | - | - | 4.00 | 424.0 |

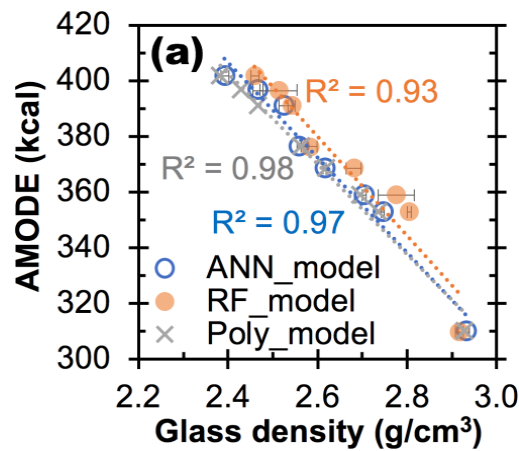

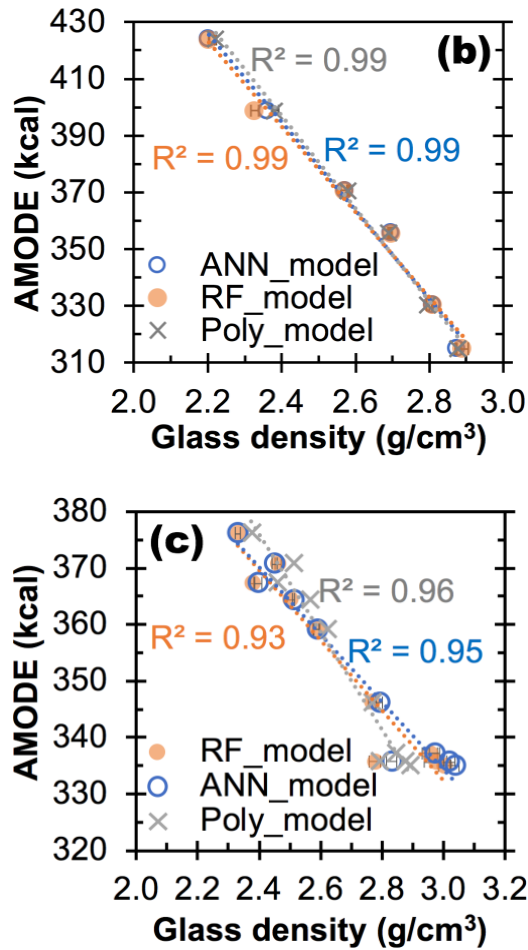

Figure 6. Correlation between predicted glass density from three density models (i.e., ANN, RF, and polynomial models) and the AMODE parameter calculated based on compositional and structural (from molecular dynamics simulations) information (see Eq. (2)). The AMODE values in (a) and (c) are obtained from ref. [33] and ref. [34], respectively.

The results in Figure 6a-c show that the densities of synthetic CAS and natural volcanic glasses are linearly and inversely correlated with their AMODE values: a higher density is associated with a lower AMODE value (and hence a higher reactivity). These inverse relationships between glass density and AMODE value ($R^2$ values for linear regression higher than 0.9 for all cases in Figure 6) are consistent with recent studies on CAS [59] and $ZrO_2$-containing soda-lime borosilicate glasses,[60] which show that glass density is inversely correlated to some structural descriptors (e.g., $F_{net}$) characterizing the network strength of glasses. Given that both the glass reactivity data (see Figure 7 and refs. [33, 34]) and density data (Figure 6) are inversely correlated with the AMODE parameter, a positive correlation between density and reactivity is expected for the glasses in the above three case studies. Regardless of the type of models used for density prediction, the $R^2$ values achieved in Figure 6 are comparable.

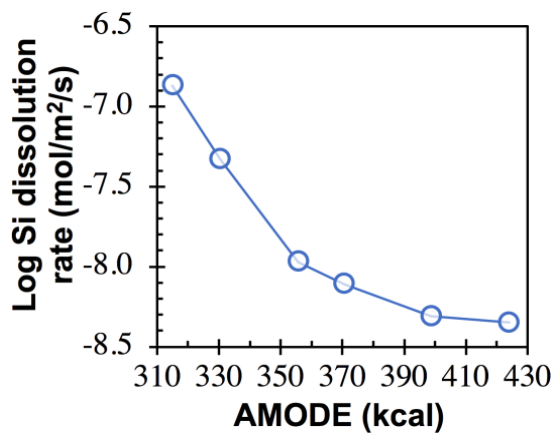

Figure 7. Comparison of the AMODE parameter of the six CAS glasses (see Table 2) and their initial glass dissolution rate at pH 13 and 20 °C (data obtained from ref. [38]). The AMODE parameter were calculated using Eq. (2).

3.4  Composition-density relationships from machine learning modeling

Using the density models from Sections 3.1 and 3.2, we predict the glass density from its chemical compositions. We demonstrate (i) to what extent density can be used as a reliable reactivity indicator and (ii) how these models can be used to explore composition-density relationships in the compositional space not (or less) covered in the database (given in Supplementary Material A). Here, we focus on CAS, MAS, and CMAS glasses, which are most relevant to the cement and concrete community, instead of covering the whole compositional space of CMASTFNKM glasses.

We first examine the impact of CaO and $Al_2O_3$ content on CAS glass density predicted from the RF, AAN, and polynomial models presented in Sections 3.2 and 3.1, with the results shown in Figures 8a, 8b, and 8c, respectively. All three models show that increasing CaO content generally leads to higher CAS glass density at all four $Al_2O_3$ levels considered. Also shown in Figure 8 are density values for $CaO$-$SiO_2$ (CS) glasses from experiments and MD simulations [63, 64] that have not been included in the original database used for model training and testing. At the same $Al_2O_3$ level, increasing CaO content (i.e., replacing $SiO_2$ with CaO) tends to lower the overall AMODE value of the CAS glass and hence increase its reactivity, since the dissociation energy of CaO (~257 kcal [44]) is much lower than that of $SiO_2$ (~424 kcal [44]). This means that the reactivity of CAS glass is positively correlated with its CaO content, which is then positively correlated with its density (as seen in Figure 8). Hence, the impact of CaO content on CAS glass reactivity should be able to be captured by its density.

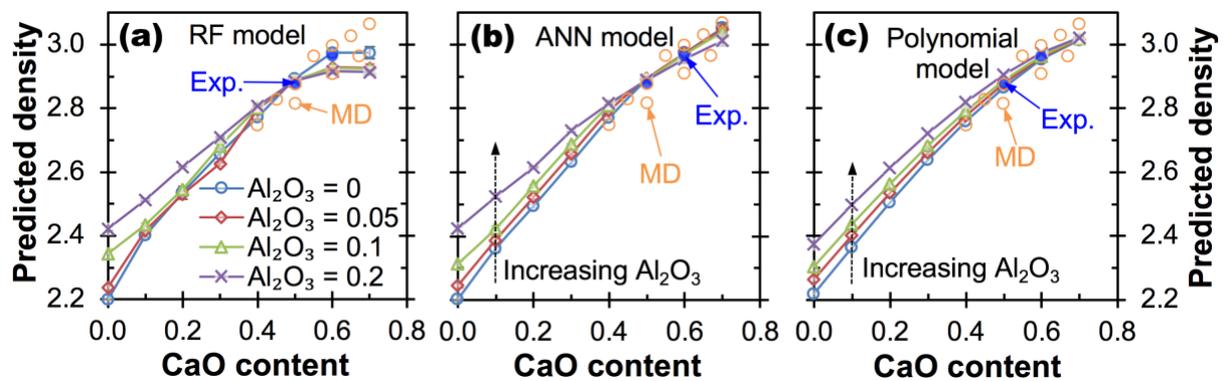

Figure 8. Impact of CaO and $Al_2O_3$ content on the density (in $g/cm^3$) of CAS glasses predicted by the (a) RF, (b) ANN, and (c) polynomial models developed in Sections 3.2 and 3.1. The error bars in (a) and (b) represent one standard deviation from twenty model predictions. Also given in the figure are density values for $CaO$-$SiO_2$ glasses from experiments (blue filled circle) and molecular dynamics (MD) simulations (orange filled circle) [63, 64] that have not been included in the original database used for model training and testing.

Furthermore, Figure 8 shows that increasing the $Al_2O_3$ level at a constant CaO content below 0.4 (i.e., replacing $SiO_2$ with $Al_2O_3$) leads to a higher density. Given that it takes less energy to break/dissociate $Al_2O_3$ (317-402 kcal [44]) than $SiO_2$ (424 kcal [44]), replacing $SiO_2$ with $Al_2O_3$ leads to a lower overall AMODE value and hence a higher reactivity. Therefore, increasing the $Al_2O_3$ level has the same positive impact on glass density and reactivity, meaning that density should be able to capture the impact of $Al_2O_3$ content on CAS glass reactivity in this compositional region. All three models also show that the extent of density increase caused by the replacement of $SiO_2$ with $Al_2O_3$ decreases with increasing CaO content. At CaO content > 0.4, the impact of $Al_2O_3$ content on CAS glass density is minimal and even reversed in the case of ML models (especially for the RF model in Figure 8a), suggesting that density becomes an unreliable reactivity indicator in this region. Given that the original database used to build the density models only contains a few data points with a CaO content above 0.6 (see Figure 1), experiments or MD simulations are needed to confirm the inverse correlation between $Al_2O_3$ content and the density predicted by RF and ANN models in this region. his, together with the observation that the smoothness of the curves in Figure 7 decreases in the order of polynomial > ANN > RF, suggests that the model transferability decreases in the order of polynomial > ANN > RF. The higher transferability of ANN than RF is consistent with the results in Figures 3 and 4, which show that the differences in the error metrics (see Table 4) between the training and testing sets are higher for the RF model than the ANN model.

Figure 9 shows the impact of MgO and $Al_2O_3$ content on MAS glass density predicted from the three density models. The RF model in Figure 9a does not capture the increasing trend of density with increasing MgO content beyond 0.5 for $MgO-SiO_2$ (MS) glasses, where data in this compositional range is not available in the original database used for model training and testing (see Figure 1 and Table S1). In particular, the predicted densities for a $Mg_2SiO_4$ glass from the RF and polynomial model are about 0.2 g/cm$^3$ lower than the corresponding experimental value (~2.90-2.92 g/cm$^3$) [65, 66]. In contrast, the ANN model in Figure 9b captures this increasing trend beyond a MgO content of 0.5, and the difference between the predicted and experimental density for a Mg$_2$SiO$_4$ glass is only about 0.05 g/cm$^3$. This result shows that the ANN model in this study exhibits higher transferability (i.e., the ability to extrapolate or generalize) than the RF model. This is consistent with the results in Figures 3-4 and Table 4, which show that the differences in the error metrics between the training and testing sets are significantly higher for the RF model than for the ANN model.

Furthermore, the general observation that increasing MgO or Al$_2$O$_3$ content in MAS glasses (i.e., replacing SiO$_2$ with MgO or Al$_2$O$_3$) leads to higher densities (Figure 9) is consistent with those seen for CAS glasses in Figure 8. Given that SiO$_2$ (~424 kcal [44]) has a higher dissociation energy than MgO (~222 kcal [44]) and Al$_2$O$_3$ (~317-402 kcal [44]), replacing SiO$_2$ with MgO or Al$_2$O$_3$ generally leads to lower AMODE values and hence higher reactivity. Therefore, the predicted density should be a reliable reactivity indicator for MAS glasses, especially when MgO content is lower than 0.4. Above 0.4, the ANN and polynomial models should still work; however, we also see that the extent of density increase caused by the replacement of SiO$_2$ with Al$_2$O$_3$ decreases slightly with increasing MgO content (Figure 9b-c). This is consistent with the observation for CAS glasses, and more research is needed to better understand this observation.

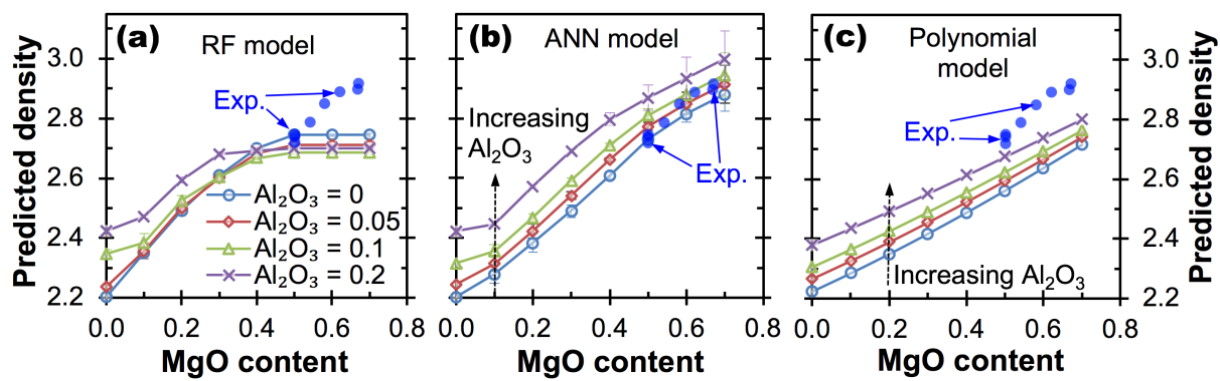

Figure 9. Impact of MgO and Al$_2$O$_3$ content on the density (in g/cm$^3$) of MAS glasses predicted by the (a) RF, (b) ANN, and (c) polynomial models presented in Sections 3.2 and 3.1. The error bars in (a) and (b) represent one standard deviation based on twenty predictions from twenty independently trained models. Also given in the figures are experimental density values for MgO-SiO$_2$ glasses [65-67] that have not been included in the original database used for training and testing the models.

Next, we evaluate the impact of MgO/(MgO+CaO) ratio and Al$_2$O$_3$ content on the predicted density of (CaO-MgO)$_{0.5}$-(Al$_2$O$_3$-SiO$_2$)$_{0.5}$ ((CM)$_{0.5}$(AS)$_{0.5}$) glasses from the RF, ANN, and polynomial models, as illustrated in Figures 10a, 10b, and 10c, respectively. We see that increasing MgO/(MgO+CaO) ratio (i.e., replacing CaO with MgO) generally leads to a lower predicted density for the (CM)$_{0.5}$(AS)$_{0.5}$ glasses. A similar impact of MgO/(MgO+CaO) on glass density has also been observed in bioactive glasses [68, 69]. The evolution curves from the ANN and polynomial models are generally smoother than those from the RF model, suggesting better transferability for the former. Comparison with the experimental density data for (CaO-MgO)$_{0.5}$-(SiO$_2$)$_{0.5}$ glasses (i.e., Al$_2$O$_3$ = 0; the data have not been used for model training and testing) in Figure 10 shows that the ANN model gives (i) a better description of the overall trend than the RF model and (ii) a more accurate density prediction (accuracy within ~0.03 g/cm$^3$, Figure 10b) than the polynomial model, especially in the Mg-rich region, where an underestimation of ~0.2 g/cm$^3$ is seen at a MgO/(MgO+CaO) ratio of 1.0 (Figure 10c). Given that CaO (~257 kcal [44]) has a higher dissociation energy than MgO (222 kcal [44]), replacing CaO with MgO (i.e., increasing MgO/(MgO+CaO) ratio) tends to reduce the overall AMODE value of the $(CM)_{0.5}(AS)_{0.5}$ glasses and hence enhance reactivity. This is consistent with a recent experimental investigation which shows that replacing 25%-50% CaO with MgO leads to slightly higher reactivity for the resulting CMAS glasses[70]. This suggests that increasing MgO/(MgO+CaO) ratio has an opposite impact on the reactivity and density of $(CM)_{0.5}(AS)_{0.5}$ glasses, i.e., positive on the former and negative on the latter, which are different from the cases of CAS and MAS glasses shown in Figures 8 and 9.

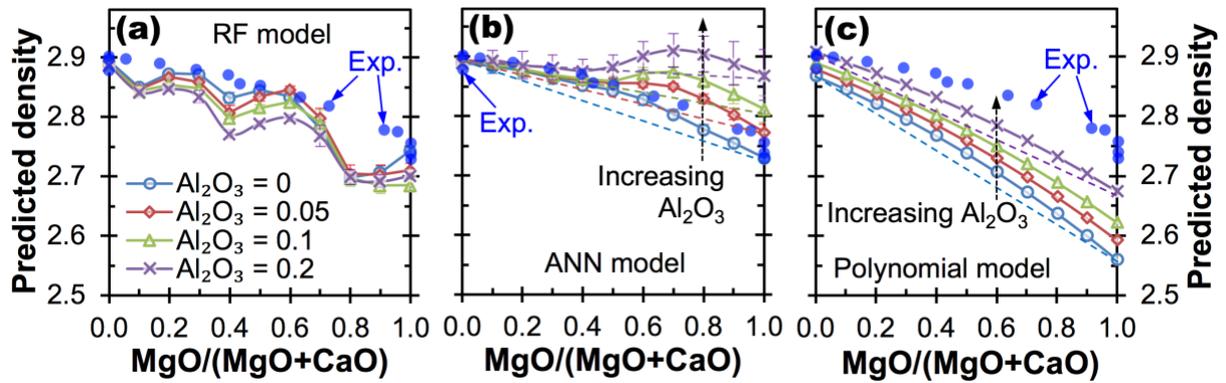

Figure 10. Impact of MgO/(MgO+CaO) and $Al_2O_3$ content on the density (in g/cm$^3$) of (CaO-MgO)$_{0.5}$-($Al_2O_3$-$SiO_2$)$_{0.5}$ (($CM)_{0.5}(AS)_{0.5}$) glasses predicted by the (a) RF, (b) ANN, and (c) polynomial models developed in Sections 3.2 and 3.1. Also given in the figures are experimental density values for $CaSiO_3$ [63], $MgSiO_3$ [66, 67] and $CaO_x$-$MgO_{(0.5-x)}$-$(SiO_2)_{0.5}$ [71] glasses that have not been included in the original database used for training and testing the models. The error bars in (a) and (b) represent one standard deviation from twenty model predictions.

In terms of the impact of $Al_2O_3$ content on density at the same MgO/(MgO+CaO) ratio, both the ANN and polynomial models show that increasing $Al_2O_3$ content (i.e., replacing $SiO_2$ with $Al_2O_3$) generally leads to a higher predicted density. The extent of increase in density due to increasing Al$_2$O$_3$ content is seen to decrease with decreasing MgO/(MgO+CaO) ratio according to the ANN and polynomial models (Figure 10b-c); however, we do not observe a similar trend for the RF model in Figure 10a. Furthermore, the ANN and polynomial models appear to suggest some mixed alkaline earth effects (i.e., the deviation from additivity in properties), where there are certain extents of deviation from linearity for the density of the glasses containing both Ca and Mg (as illustrated by the dashed lines in Figure 10b-c). These mixed alkaline earth effects in silicate glasses have been widely observed for different properties, including density, Young's modulus, passion ratio, toughness, hardness, the activation energy for diffusion, and glass transition temperature [72-74]. The experimental density data in Figure 10 for CaO$_x$-MgO$_{(0.5-x)}$-(SiO$_2$)$_{0.5}$ glasses [71] also exhibit a certain level of deviation from linearity (i.e., mixed alkaline earth effect). However, some studies on (CaO)$_{(0.08-x)}$-(MgO)$_x$-(Na$_2$O)$_{0.16}$-(Al$_2$O$_3$)$_{0.16}$-(SiO$_2$)$_{0.6}$ [73] and (Na$_2$O)$_{0.2}$-(CaO)$_{(0.1-x)}$-(BaO)$_x$-(SiO$_2$)$_{0.7}$ [75] glasses showed no obvious mixed alkaline earth effects for density, which could be related to the relatively low alkaline earth content in these studies (< ~10 mol. %) compared with those in Figure 10 and ref. [71].

### 3.5 Broader impact and limitations

*3.5.1 Broader impact*

This study has demonstrated that data-driven ML models exhibit a superior ability to predict glass density for CMASTFNKM glasses compared with empirical density models based on ionic packing ratio. In particular, a single layer ANN model exhibits some level of transferability to a compositional space not (or less) explored during the training and testing, which is consistent with our recent study on data-driven prediction of quartz dissolution rates [43]. This is encouraging since the model can then be used to probe wider compositional spaces that are less explored in the literature, either experimentally or computationally. Furthermore, this study also explicitly shows that density can be used as a reliable indicator of reactivity for certain aluminosilicate glasses (e.g., CAS and volcanic glasses) in certain compositional spaces. This is potentially important for the cement community because density is a commonly measured property for the glassy low-$CO_2$ raw materials (e.g., blast-furnace slag and fly ash) used to lower the $CO_2$ footprint of cement and concrete and hence could be potentially used as a screen to evaluate the reactivity of these glassy materials. One potential challenge here is that these low-$CO_2$ raw materials often contain a mixture of crystalline and amorphous phases, rendering it difficult to quantify and estimate the density of the glassy phases (the main reactive components) present in these materials. One potential remedy to this challenge is to combine these data-driven density models with phase quantification experiments, which include (i) SEM-EDS-based analysis that allows both amorphous phase quantification and spatially-resolved compositional analysis [76], (ii) the Partial Or No Known Crystal Structure (PONKCS) approach based on X-ray (or neutron) diffraction [77], and (iii) X-ray pair distribution function analysis combined with MD simulations [78]. These combinations may help us determine not only the amorphous content but also the chemical composition and density of the amorphous phases and hence may enhance our understanding of the reactivity of the compositionally and mineralogically complex low-$CO_2$ waste materials, which is critical to their recycling for low-$CO_2$ concrete production. In fact, one recent study shows that a data-driven approach exhibits great promise for predicting the content of the glassy phase as well as its chemical composition in fly ash [79]. Finally, these methods and analysis could be potentially extended to study other glass systems and the associated composition-property and density-property relationships.

*3.5.2 Limitations*

The reliability of density as a reactivity indicator across a broader compositional space needs to be evaluated. More importantly, the reactivity of silicate glasses is highly complex, and, in addition to the glass composition and structure touched upon in this investigation, reactivity can also be influenced by solution chemistry, pH, temperature, particle size distribution, degree of amorphousness (and the presence of crystalline phases), and potential phase segregation [36, 80]. The last two factors are especially relevant to the amorphous aluminosilicates used for cement and concrete production (e.g., fly ash and blast furnace slag) because the presence of crystalline phases and phase segregation have been widely reported, especially for fly ash [30, 76, 81]. In particular, the reactivity of CMAS glass can be even more complex. On the one hand, electronic density state calculations near the Fermi level show that Ca sites are more reactive than Mg sites [48], which is consistent with the general observation in mineral dissolution that Ca-based silicates (e.g., $Ca_2SiO_4$) have a much higher dissolution rate than corresponding Mg-based silicates (e.g., $Mg_2SiO_4$) [82]. On the other hand, Mg (as compared to Ca) promotes the formation of free oxygen sites, which are the most reactive sites in CMAS glasses [48, 83]. Furthermore, there are significantly fewer data points in the database for glasses containing high levels of MgO and CaO; as a result, the associated density prediction and composition-density relationship in this region requires more investigation. Hence, care is needed when using density as a reactivity indicator for $(CM)_{0.5}(AS)_{0.5}$ glasses. Also, there are significantly fewer data points for glasses containing MnO, FeO, $Fe_2O_3$, and $TiO_2$ (as seen in Figure 1 and Table S1), rendering it difficult to explore the composition-density relationships in these compositional subspaces. As more glass density data is needed to build more robust and insightful ML models, high-throughput atomistic modeling (e.g., MD simulations) provides a promising direction of further study [23].

## 4 Conclusions

Here, a room-temperature density database has been constructed for glasses in the compositional space of $CaO$-$MgO$-$Al_2O_3$-$SiO_2$-$TiO_2$-$FeO$-$Fe_2O_3$-$Na_2O$-$K_2O$-$MnO$ (CMASTFNKM) based on data extracted from ~140 literature studies. Based on this database, we have developed several models for density prediction, including linear and second-order polynomial models based on ionic packing ratio and two types of machine learning models (i.e., random forest (RF) and artificial neural network (ANN)). We first showed that both the linear and polynomial models give significantly better density prediction (MAPE = ~1.40-1.57 %; $R^2$ values = ~0.88-0.91) than previously developed models using a constant ionic packing ratio (MAPE = ~2.78-4.64 %; $R^2$ values = ~0.28-0.75). We then showed that both the RF and single layer ANN models exhibit even higher predictive performance, reducing (increasing) the MAPE ($R^2$ values) for the 15% testing set to ~0.59-0.79 % (~0.96-0.98). The potential of using density as a reactivity indicator was then evaluated for fourteen synthetic $CaO-Al_2O_3-SiO_2$ (CAS) and seventeen natural volcanic glasses reported in three high-quality literature studies. The results show that the predicted densities from all three models (i.e., polynomial, RF, and ANN) capture well the relative reactivity of the different glasses in each study (especially for the simpler CAS glasses), with a higher density associated with a higher reactivity. This positive correlation between glass density and reactivity ($R^2$ values of 0.73-0.97 for linear regression) is attributed to the inverse correlation between glass density and the average metal-oxygen dissociation energy (AMODE) parameter of the glasses (with $R^2$ values higher than 0.9 for linear regression), where glasses with a lower AMODE value require less energy to break all the metal-oxygen bonds and hence tend to be more reactive. The AMODE values for one of the case studies were calculated from molecular dynamics (MD) simulations performed here, while those for the other two case studies were obtained from the literature.

Finally, we compared the composition-density relationships from three model predictions (i.e., RF, ANN, and polynomial) for selected compositional subspace (i.e., CAS, $MgO-Al_2O_3-SiO_2$ (MAS) and $(CaO-MgO)_{0.5}-(Al_2O_3-SiO_2)_{0.5}$ $((CM)_{0.5}(AS)_{0.5})$ glasses). The comparisons generally show that the ANN model exhibits (i) better transferability (i.e., ability to extrapolate to compositional space not (or less) covered by the original database) than the RF model and

(ii) more accurate density predictions than the polynomial model for these glasses. The results also suggest that density should generally be a reliable reactivity indicator for CAS and MAS glasses, capturing the impact of CaO (or MgO) and $Al_2O_3$ on reactivity reasonably well, especially when CaO (or MgO) < 0.4. However, it is less reliable for the $(CM)_{0.5}(AS)_{0.5}$ glasses considered here, with the sensitivity of density to compositional change varying considerably depending on the type of model used for density prediction, MgO/(MgO+CaO) ratio, and $Al_2O_3$ content. The ANN and polynomial models also predict several interesting observations in density-composition relationships, including (i) a reduction in the extent of density increase induced by replacing $SiO_2$ with $Al_2O_3$ at increasing CaO or MgO level or decreasing MgO/(MgO+CaO) ratio, and (ii) mixed alkaline earth effect (deviation from linearity) for $(CM)_{0.5}(AS)_{0.5}$ glass density. While the mixed alkaline earth effects have been widely observed in the literature, the first observation still needs to be verified with experiments or MD simulations.

# 5   Acknowledgment

This work is supported by funding from the MIT-IBM Watson AI Lab. The MD simulations were performed using computational resources in the Microsystems Technology Laboratories at MIT. The authors also acknowledge Dr. Zach Jensen and Mr. Christopher Karpovich for their suggestions on ML modeling.

# 6   Supplementary Materials

6.1   Supplementary Material A

6.2   Supplementary Material B

1. Database statistics;

2. Optimization of hyperparameters;

3. Calculation of error metrics;

4. Single metal-oxygen bond strength parameter;

5. Estimation of the CMAS glass density at different temperatures;

6. Ionic packing factor;

7. Predictive performance of constant ionic packing factor;